# *A Study on Lamprey's Population Based on Sex-Ratio-Related Growth-Balance Model*

**Zuhua Ji, Jiarui Chen, Zihang Wang**

*School of Physical and Mathematical Science, Nanjing Tech University, Nanjing, 210000, China*

*Abstract:* Lampreys are one of the oldest species in the world, living longer than dinosaurs, which is related to the ability to change the sex ratio during their lifespan. In this paper, to understand how sex ratio and food quantity affect the population growth rate of lampreys, the researchers draw inspiration from the logistics model and established a model called EcoSexChange(ESC), which results in a population initially increasing and then stabilizing, a reasonable outcome that may apply to other organisms with significant differences in consumption between sexes. Subsequently, this paper develops the Sex Ratio Adaptation Eco Impact (SRAEI) model based on the ESC model using the ABM algorithm to simulate how the population of lampreys, whose lives are divided into seven stages, grows and stabilizes. Then introduces a sudden disaster factor in the middle of the simulation, while also comparing lampreys that cannot adjust their sex ratio. The results of this paper are of great reference significance for people to analyze the population changes of lampreys in different living environments, and they are also easy to apply to other species with large differences between males and females.

## 1. Introduction

Lampreys are among the oldest species in the world, having existed for 360 million years. This species is elongated, with seven pairs of gills. Research has revealed that through evolution, lampreys can change their gender during maturation, which is one of the reasons they have not gone extinct. Lampreys play a complex role in the ecosystem. In some places, they are an important part of the ecosystem, and in other places, they become the destroyers of ecological balance, such as the Great Lakes region of North America [1].

The average lifespan of lampreys is 7 years. They spend the first four years growing in freshwater, undergo metamorphosis in the fifth year to transition to the sea, and in the seventh year, they migrate to spawn. Another feature of lampreys is that their sex is not determined at birth. Instead, it is determined at maturity based on the amount of nutrients they absorb. Those that absorb more nutrients will become larger females, while those that absorb less will become smaller males. Because of their high reproductive capacity, they caused significant disruption in the Great Lakes ecosystem. Many factors can lead to the sex change in lampreys. Hence, understanding the effects of lampreys' metamorphic abilities on themselves and the environment is a matter of urgency.

Yantao Luo once used a parasitic and predation model that included time delay factors to predict the population growth of birds and mammals [2], but a disadvantage is that the impact of the gender ratio factor was too much overlooked. Bensch H, Thorley J, and Finn K incorporated the factors of gender ratio and food quantity when studying the population of rats [3], but the downside is that the model was overly complex.

The advantage of the model proposed in our article lies in its ability to integrate the impact of gender ratio and food quantity into a simple logistic model, while also being relatively easy to use. Moreover, the model itself is not highly dependent on data; it only requires fitting a few parameters with data to conveniently arrive at reasonable conclusions.

## 2. The model of sex ratio adjustment in Lampreys

### 2.1 The establishment of the EcoSexChange model

We collected data about sea lampreys in the Great Lakes. The ratios of males in the southeastern United States vary from 9% to 49%, and the ratio of males becomes 78% when the sea lampreys live in the area of colder regions [4]. Based on the findings, we managed to establish a new model called ESC to talk about all sex ratios of the sea lampreys.

For the first part of the model, we accept an assumption: The sea lampreys population growth directly depends on the proportion of female pregnant sea lampreys. So the growth ratio of population g is proportional to the scale factor k, conception rate among females $R_p$ and Female R. According to this assumption, we can get the growing rate g as follows,

$$g = kRR_p \tag{1}$$

where Rp is the conception rate among females, R is the female rate among all the lampreys, and k is the proportionality constant. We can find that: the growing rate g as follows,

$$R_p = s(1 - R) \tag{2}$$

If we see the fertility index s as a constant, we can take equation 2 into equation 1, and get the new equation:

$$g = ksR(1 - R) \tag{3}$$

The equation represents a symmetric parabola over the interval from 0 to 1.

Because both the male and female sea lampreys will die in a few days after maturity, this is what we conclude: A female sea lamprey can only be impregnated once in her lifetime, while male sea lampreys will mate more or less with female sea lampreys based on the gender ratio. As a result, s should not be a constant. We do some corrections on s related to R.

We defined -k as the mating ratio, and use $s_0$ as a constant. $s_0$ is the average fertility index in the lampreys population.

$$s = -k'(1 - R) + s_0 \tag{4}$$

Take the equation 4 into equation 1. It can be simplified to a new equation:

$$g = k[-R^3 + (1 + k' - s_0)R^2 + (s_0 - k')R] \tag{5}$$

We defined a new character m to simply the equation

$$s_0 - k' = m \tag{6}$$

Moreover, we introduce Net Growth Rate to make the equation more simplified. k is no connection between g and R. As k > 0, the tendencies of the residual equation are similar to the equation 5, and then we get a new simplified expression:

$$-R^3 + (1-m)R^2 + mR \qquad (7)$$

Based on the previous circumstances and the boundary conditions

$$k > 0,\ k' > 0,\ s_0 > 0,\ m \geq 0 \qquad (8)$$

Differentiating the above equation 7, we obtain

$$-3R^2 + 2(1-m)R + m \qquad (9)$$

Obviously, we can learn that when R = 0 and R = 0.5, the expression 10 is greater than 0. When R = 1, the expression is less than 0.

We can find that if food decreases, the female ratio will decrease slower.

This model represents changes in the population growth rate when considering only the influence of gender ratio. We can observe that when not considering environmental factors, due to the fact that male sea lampreys can mate with more female sea lampreys than their own number, the population growth rate is fastest when there are slightly more female sea lampreys than males. However, when considering only the impact of the environment on the population, as female sea lampreys rely more on a favorable environment, including more food [5], and their large bodies are more vulnerable to predation, the environmental requirement $P_f$ for a female sea lamprey is significantly greater than the environmental requirement $P_m$ for a male sea lamprey.

When the environmental carrying capacity remains constant, we can get

$$SRP_f + S(1-R)P_m = E \qquad (10)$$

where S is the total population of the lampreys.

Simplify to obtain

$$E = S[R(P_f - P_m) + P_m] \qquad (11)$$

At this point, when we substitute it along with equation 7 into the population quantity equation 13, we obtain a partial differential equation model for total number of the lampreys S with respect to the two independent variables female ratio R and time t. If we calculate it directly, it becomes quite complex. Therefore, we use a computer to discretize it and employ the finite difference method for numerical solution. We select the following parameters to generate the graphs.

In this model, we can observe that when the environmental carrying capacity K remains constant, assuming that the population hasn't yet reached the vicinity of the carrying capacity $P_m$, they may live with their original gender ratio. However, as the population gradually approaches to K, their R will start to change, moving in the direction indicated in the graph towards a new value. Eventually, it stabilizes near the value most suitable for the population's survival.

When considering the relationship between the lampreys and other species in the larger ecosystem, we can simplify the proportionality coefficients in the equation to constants to obtain a model.

We talked about that how the f/m ratio influences the population above, now we are measuring how much food they will consume on the base of that female lampreys need more food to consume than male ones.

We use $f\left(\frac{\partial S}{\partial R}\right)$ to measure the death rate because the death rate and R is connected to the contribution of the S growth.

$$\frac{\partial S}{\partial t} = k\left\{[-R^3 + (1-m)R^2 + mR] - f\left(\frac{\partial S}{\partial R}\right)E\right\}S \qquad (12)$$

In order that this equation can be solved in a simpler way, we simplify it, specifically by setting $f\left(\frac{\partial S}{\partial R}\right)$ to a constant u.

Substituting equation 11, we get

$$\frac{\partial S}{\partial t} = k\{[-R^3 + (1-m)R^2 + mR] - u[SRP_f + S(1-R)P_m]\}S \qquad (13)$$

## 2.2 The ABM algorithm simulation based on ESC model

We modify the Agent-Based Modeling (ABM) to get our new SRAEI model. ABM uses the computer simulation, predicting population dynamics from individual lamprey behavior. First, we defined our Agent is sea lamprey, which is able to act autonomously and is able to interact with other Agents and the environment.

Then we decide what the sea lampreys can do, then using the computer simulation to get these data.

We also compare them with a hypothetical group of lampreys that could not change their sex ratio based on food quantity, so that their advantages and disadvantages can be clearly realized. Figure 1 is the idea behind the algorithm.

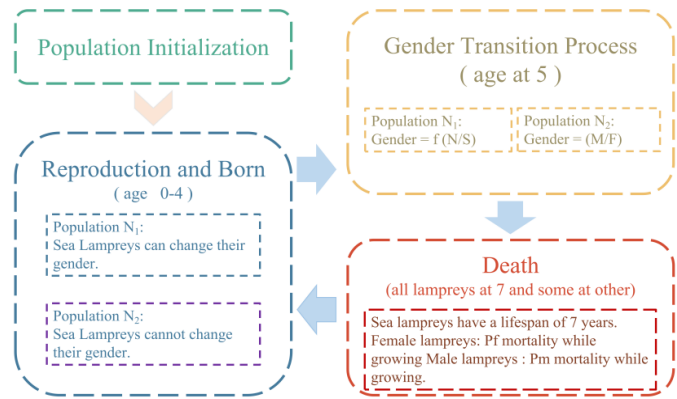

Figure 1: The ABM algorithm implementation of the SRAE model

Using single variable approach, we test two population of sea lampreys, called population $N_1$ and population $N_2$. $N_1$ have the ability to change their gender at 5 years old according to the environment, while $N_2$ cannot change their gender as the environment change.

In the $N_2$ population, the food consume is equal between the females and the males. In the $N_1$ population, the food consume is different, females need more food to gain reproduction.

We first initialize the population of sea lampreys, setting a suitable number at about 500. Assuming that the sea lampreys need three stages to go through: Reproduction and Born, Gender Transition Process and Death, we can simply the SexRatioAdaptationEcoImpact model.

We divided the life of a lamprey into seven stages, with the length of a stage being a unit of time, modeling how a group of lampreys in different stages would change their population over time. When sea lampreys are at fifth stage, sea lampreys will change their gender according to the environment, while the female are more likely to die because of their size, being catched by people and so on. As a result, we assume each lamprey will die at 7 years old, as well as male lampreys has the possibility of $P_m$ die, while female lampreys has the possibility of $P_f$ die.

In order to test the rate of population recovery, we set a disaster and food decreasing after the population of $N_1$ and $N_2$ become stabilize, and then we can see what will happen to them.

The most important part of this algorithm is how to quantify the relationship between individual mortality and food intake. This can be done using the conclusions of the previous model.

From the stability theory of differential equations, it is known that there is an environmentally permissible maximum population, or we can say environmental carrying capacity, at its stabilization point obtained from equation 14.

$$S_{max} = \frac{-R^3 + (1-m)R^2 + mR}{[RP_f + (1-R)R_m]u} \quad (14)$$

In this model we use $2 \times \left(1 - \frac{S}{S_{max}}\right)$ to describe the survival ratio. The theoretical maximum survival ratio is $S_{max}$, and the practical survival ratio decides S cannot reach $S_{max}$. According to the equation 14, we can get the death rate D:

$$D = \frac{1}{\frac{2S(RQ+1)}{-R^3 + (1-m)R^2 + mR} - 1} \quad (15)$$

Up to this point, we have obtained all the parameter conditions for conducting ABM algorithm simulations and have established a connection with the population's sex ratio.

The remaining proportion constants can be obtained by parameter fitting in practical problems. They will only affect the magnitude of the specific values, but will not affect the trend of the image changes. We can then examine the impact of this sex ratio adjustment mechanism on the lamprey population.

## 3. Model solution results

### 3.1 The results of ESC model

If we plot the function of Equation 7 as shown in the figure, we can see that, in the case where a female can only mate once in her lifetime, the natural growth of the population is always fastest when the female ratio is slightly more than half. Moreover, as m approaches 1, the graph gradually tends to become symmetrical from left to right, as shown in Figure 2.

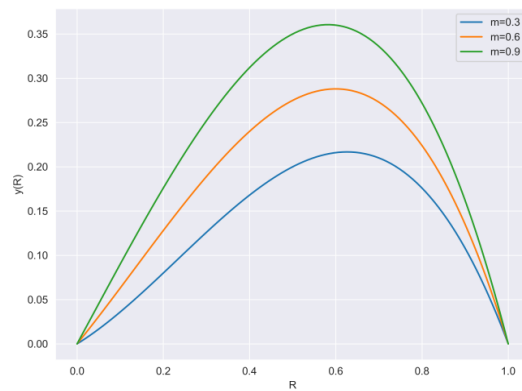

Figure 2: The growth rate changes with R under different m

The figure shows that the population growth rate decreases when the number of females is too high or too low. When food is not considered, the population growth rate is the fastest when the proportion of females in the population is slightly higher than half. This is because a female can only mate once, while a male can mate multiple times. The lower the parameter m, the stronger the male mating ability, and the higher the female proportion at the maximum growth rate.

After selecting appropriate parameters and solving differential equation 14, we obtain a binary function of the sex ratio R and time t. Observing this function, we can find that as the population increases and food decreases, the original "growth type" with a slightly higher female ratio than male turns into a "balanced type" where the male ratio is significantly higher than the female, What makes this special is that it is an active change.

This result can be seen from Figure 3, which uses the parameter that the female food consumption is 5 times that of the male, the food amount is constant, and for the lamprey population, it is in a state of food abundance at the beginning, and then the food gradually becomes insufficient. Here, the change is natural. We will see about unnatural changes in disaster simulations.

As can be seen, when the final population reaches equilibrium, the proportion of females will drop to around 20% to 30%. Of course, the resource consumption here can be not only limited to food, but also the safety requirements of the surrounding environment, or the temperature, humidity and other requirements. In many aspects, the requirements of female lampreys are higher than those of males.

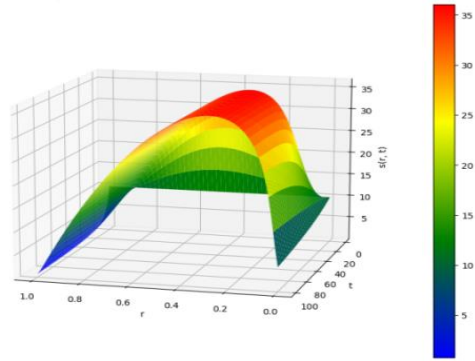

Figure 3: The solution results of the ESC model (3D)

Moreover, by looking at the graph of the cubic function, we can also discover that when the amount of food decreases, if the rate of growth decreases in the direction of reducing the number of females, its decline is slower than in the direction of increasing females. This way, the population can always maintain itself as large as possible at any time, further verifying the scientific nature of the ESC model.

Due to the greater hunting ability of female lampreys compared to males, different sex ratios ultimately reflect on the overall predation efficiency of the lamprey pack. This impact can be intuitively observed using the simple Lotka-Volterra model.

Through the Lotka-Volterra model, we can get the equations below:

$$\frac{dS}{dt} = m_1 S - DS + QbSP \quad (16)$$

$$\frac{dP}{dt} = m_2 S - bPS \quad (17)$$

where D is the Natural Mortality Rate, Q is the Population Size of Prey, and b is used as a correction factor.

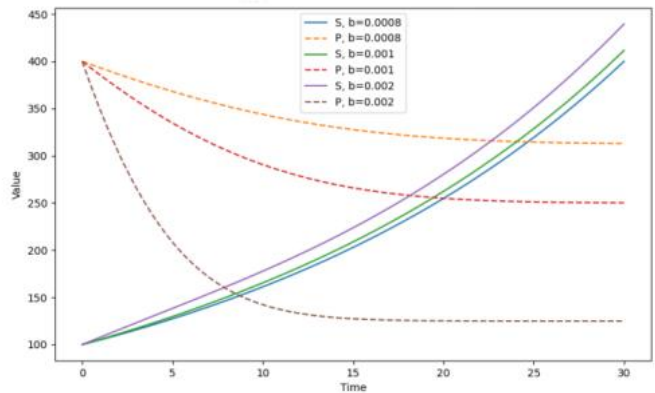

Figure 4: The impact of different female ratios on prey populations

As shown in Figure 4, the direct result of an increase in females is that the population of fish species that are preyed upon by lampreys will decrease faster, and even have the risk of extinction.

## 3.2 The results of SRAEI model

In the results of the SRAEI model, we can also see outcomes that are more similar to real situations. That is the Figure 5. In addition to the difference in quantity, the images also show the difference in the sex ratio of the two populations.

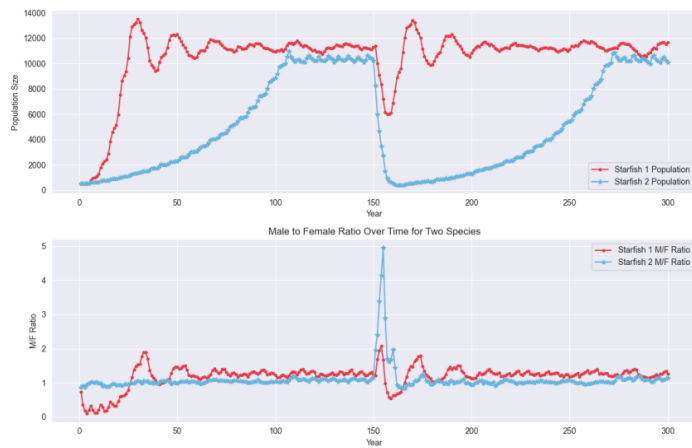

Figure 5: The comparative simulation results of the SRAE model

In Figure 5, the red line represents population $N_1$, while the blue line represents population $N_2$. Compared to populations that cannot adjust their sex ratio, those that can adjust their sex ratio grow faster during growth phases, have larger populations at equilibrium and have stronger recovery capabilities after sudden disasters. These are all advantages manifested by this regulatory mechanism. However, we can also observe that this type of population experiences fluctuations in numbers at the moment of reaching equilibrium, which will lead to an even greater waste of already scarce food resources during this phase. In contrast, populations that cannot adjust their sex ratio waste less, or under the conditions of the logistic model, almost no waste occurs. This is their disadvantage. In areas suffering from lamprey invasions in recent years [6], our work may provide some reference for prevention and control.

## 4. Conclusions

In this article, we successfully established the ESC model and the SRAEI model. In the use of these two models, only a few parameters need to be extracted from the existing data to obtain very accurate and reasonable conclusions, which can provide a basis for predicting the future development of the population. The introduction of the gender ratio factor in our models is successful and worth referencing. Through the ESC model, we obtained the conclusion that more female lampreys increase the consumption of food by the entire population, which increases predation on other species in the larger ecosystem, and their population grows more rapidly.

However, there are still areas where these two models can be improved. First, we cannot determine the process of changes in the gender ratio over time, only the results of the changes. This changing process may require a more complex model to resolve, as it may involve the impact of factors within a broader range, such as ecosystems. Second, during the simulation process, we simplified the seven stages from birth to death of lamprey to be of equal duration. Although this is reasonable in terms of the results, if the problem involves more precise quantification and the impact of other organisms on the lamprey population, it may be necessary to further refine the specific duration of each stage. Therefore, integrating the impact of ecosystems for more precise quantitative research may be a significant direction for the future.